# Enhancement of Fusion Reactivity under Non-Maxwellian Distributions: Effects of Drift-Ring-Beam, Slowing-Down, and Kappa Super-Thermal Distributions

Haozhe Kong[1,2,3], Huasheng Xie[2,3,*], Bing Liu[2,3], Muzhi Tan[2,3], Di Luo[2,3], Zhi Li[2,3], and Jizhong Sun[1,*]

[1]School of Physics, Dalian University of Technology, Dalian 116024, China
[2]Hebei Key Laboratory of Compact Fusion, Langfang 065001, China
[3]ENN Science and Technology Development Co., Ltd., Langfang 065001, China

E-mail: xiehuasheng@enn.cn and jsun@dlut.edu.cn

**Abstract**

Non-Maxwellian distributions of particles are commonly observed in fusion studies, especially for magnetic confinement fusion plasmas. The particle distribution has a direct effect on fusion reactivity, which is the focus of this study. We investigate the effects of three types of non-Maxwellian distributions, namely drift-ring-beam, slowing-down, and kappa super-thermal distributions, on the fusion reactivities of D-T (Deuterium-Trillium) and p-B11 (proton-Boron) using a newly developed program, where the enhancement of fusion reactivity relative to the Maxwellian distribution is computed while keeping the total kinetic energy constant. The calculation results show that for the temperature ranges of interest to us, namely 5-50 keV for D-T and 100-500 keV for p-B11, these non-Maxwellian distributions can enhance the fusion reactivities. In the case of the drift-ring-beam distribution, the enhancement factors for both reactions are affected by the perpendicular ring beam velocity, leading to decreased enhancement in low temperature range and increased enhancement in high temperature range. However, this effect is favorable for p-B11 fusion reaction and unfavorable for D-T fusion reaction. In the slowing-down distribution, the birth speed plays a crucial role in both reactions, and increasing birth speed leads to a shift in the enhancement ranges towards lower temperatures, which is beneficial for both reactions. Finally, the kappa super-thermal distribution results in a relatively large enhancement in the low temperature range with a small high energy power-law index $\kappa$. Overall, this study provides insight into the effects of non-Maxwellian distributions on fusion reactivity and highlights potential opportunities for enhancing fusion efficiency.

## 1.Introduction

In fusion plasmas, the velocity distribution of fusion reactants frequently deviates from the Maxwellian distribution due to various factors, such as heating and fusion high-energy products. For example, recent experiments at the National Ignition Facility (NIF) have provided evidence for suprathermal ion distribution in burning plasmas[1]. The fusion yields depend on the distribution functions of the reactants. Both experimental[2] and theoretical[3-10] evidence suggest that non-Maxwellian distributions can have a beneficial effect on the fusion reaction, which is a promising indication for relaxing [7] the Lawson criterion[11], particularly for p-B11 fusion reaction. Previous research on fusion reactivity with non-Maxwellian distributions has primarily assumed the reactant distribution to be approximately variants of the Maxwellian distribution, such as bi-Maxwellian[4,5] distribution, drift tri-Maxwellian[3] distribution, and drift bi-Maxwellian[7] distribution. Moreover, ongoing research is exploring other types of non-Maxwellian distributions, such as the Dagum distribution[12]. In 2006, Nakamura et al. [10] confirmed that non-Maxwellian knock-on perturbations caused by close collisions between alpha particles and background ions would result in reactivity enhancement in fusion reactions. Subsequently, studies on the effect of reactant distributions with high-energy tails in fusion reactions, such as the kappa distribution[12,13], have been conducted. Overall, these findings suggest that non-Maxwellian distributions have a significant impact on fusion reactivity and open up new avenues for enhancing fusion efficiency.

The fusion reactivity in the study mentioned above can be expressed as:
$$\langle\sigma v\rangle = \iint d\boldsymbol{v}_1 d\boldsymbol{v}_2 \sigma(|\boldsymbol{v}_1 - \boldsymbol{v}_2|)|\boldsymbol{v}_1 - \boldsymbol{v}_2| f_1(\boldsymbol{v}_1) f_2(\boldsymbol{v}_2), \quad (1)$$
where $f_j(\boldsymbol{v}_j)$ are normalized velocity distribution functions, i.e., $\int f_j(\boldsymbol{v}_j) d\boldsymbol{v}_j = 1$ with $j = 1,2$ representing reactant species, respectively, $\sigma$ represents the fusion cross section, and $d\boldsymbol{v}_j = dv_{jx} dv_{jy} dv_{jz}$.

Equation (1) presents a six-dimensional integral in velocity space, making analytical solutions difficult to obtain. However, when considering the Maxwellian[14-16] distribution and some of its variants[3-5,7,8,17-20], lower-dimensional integral forms can be derived. The analytical work on these simplified forms has provided valuable insights into fusion reactivity. For instance, studies on Maxwellian distributions with drift have explored the impact of drift on fusion reactivity[3,7]. Moreover, research on anisotropic Maxwellian distributions has revealed that anisotropy can enhance fusion reactivity in the low temperature range [3-5,7]. Additionally, studies on Maxwellian distributions with different reactant temperatures have shown that lighter reactants with higher temperature relative to heavier ones tend to promote fusion reactivity[13]. A summary and review of the low-dimensional integral forms mentioned above can be found in references[7]. Pursuing analytical solutions is useful for two reasons: computational efficiency when dealing with low-dimensional integrals and a better understanding of the physics in a simpler analytical form[4,5,7].

However, while analytical methods have yielded valuable insights into the behavior of Maxwellian distributions and some of their variants, they often prove insufficient when dealing with more complex non-Maxwellian distribution forms. Consequently, numerical calculation of Equation (1) becomes essential for a more in-depth study of the complex distribution functions that have proven difficult to handle analytically, such as the commonly observed drift-ring-beam distribution, slowing-down distribution, and kappa super-thermal distribution in fusion plasmas. In this study, we have developed a program to numerically calculate the enhancement in fusion reactions for these distributions, as well as for the Maxwellian distribution and its variants. We conduct a detailed analysis to examine how the drift-ring-beam distribution, slowing-down distribution, and kappa super-thermal distribution affect the reactivity in D-T and p-B11 fusion reactions, providing new insights into the behavior of these complex plasma systems.

In Section 2, we provide a comprehensive description of the model utilized in this study. Section 3 focuses on investigating the effects of key parameters associated with the drift-ring-beam distribution, slowing-down distribution, and kappa super-thermal distribution on the enhancement of fusion reactions. Our analysis aims to provide a detailed understanding of the complex distribution functions and their impact on fusion reactivity in D-T and p-B11 fusion reactions. Finally, in Section 4, we present a summary of our findings.

## 2. Model and strategy

In fusion experiments, non-Maxwellian distribution functions, such as the drift-ring-beam, slowing-down, and kappa super-thermal distributions, are commonly observed in reactants. Hence, the main focus of this study is to analyze and understand the effects of these distributions. The drift-ring-beam distribution, which is of particular interest in this study, has the following mathematical form [21,22]:
$$f(\boldsymbol{v}) = \frac{1}{\pi^{3/2} A v_{t\parallel} v_{t\perp}^2} \exp\left(-\frac{(v_\parallel - v_{dz})^2}{v_{tz}^2}\right) \exp\left(-\frac{\left(\sqrt{(v_x - v_{dx})^2 + (v_y - v_{dy})^2} - v_{dr}\right)^2}{v_{t\perp}^2}\right), \quad (2)$$
where $\int f(\boldsymbol{v}) d\boldsymbol{v} = 1$, $A = \exp\left(-\frac{v_{dr}^2}{v_{t\perp}^2}\right) + \frac{\pi^{1/2} v_{dr}}{v_{t\perp}} \text{erfc}\left(-\frac{v_{dr}}{v_{t\perp}}\right)$, erfc(x) is the complementary error function, $v_{t\parallel}$ and $v_{t\perp}$ are the parallel and perpendicular thermal velocities, $v_{dx}$, $v_{dy}$, $v_{dz}$ are the drift velocities in $x$ (perpendicular 1), $y$ (perpendicular 2) and parallel directions, respectively, and $v_{dr}$ is the perpendicular ring beam velocity. The slowing-down distribution is as follows[22]
$$f(\boldsymbol{v}) = \frac{3}{4\pi \ln[1 + v_b^3/v_c^3]} \frac{H(v_b - v)}{v^3 + v_c^3}, \quad (3)$$
where $v^2 = v_x^2 + v_y^2 + v_z^2$, $v_b$ is the birth speed, $v_c$ is crossover speed, H(x) is the Heaviside function. The kappa super-thermal distribution is[13,23]
$$f(\boldsymbol{v}) = \frac{1}{(\pi \kappa v_{th}^2)^{3/2}} \frac{\Gamma(\kappa + 1)}{\Gamma(\kappa - 1/2)} \left(1 + \frac{v^2}{\kappa v_{th}^2}\right)^{-(\kappa + 1)}, \quad (4)$$

where $v_{th}$ is an equivalent thermal speed, $\Gamma(x)$ represents the gamma function, $\kappa$ is a parameter which determines high energy power-law index. It is to be noted that for $\kappa \to \infty$, kappa super-thermal distribution tends to Maxwellian distribution.

As previously mentioned, analytical solutions for the low-dimensional integrals of fusion reactivity for the three aforementioned distributions pose significant challenges. Hence, numerical calculations have emerged as a more practical and feasible approach to investigating their effects on fusion reactivity. Our research focuses on the impact of key parameters of the reactant distribution on the enhancement of fusion reactivity relative to the Maxwellian distribution while keeping the total kinetic energy constant. To solve the total energy equation numerically, we utilize a fixed parameter ratio. In the total energy equation, we assume that reactant 2 (heavier reactant) follows the Maxwellian distribution, while the distribution of reactant 1 (lighter reactant) is either the drift-ring-beam, slowing-down, or kappa super-thermal distributions. Next, we provide a detailed overview of our model and strategy.

Firstly, a numerical calculation is applied to the energy equation, in which the total energy $E_{\text{total}}$ and parameter ratios are held constant, in order to derive precise values for the parameters. The energy equation is written as

$$E_k = E_{\text{total}}, \quad (5)$$

where $E_k = 0.5 \sum_{j=1}^{2} m_j \int v_j^2 d\boldsymbol{v}_j f_j(\boldsymbol{v}_j)$ represents the total energy of the two reactants. For the drift-ring-beam distribution, $E_k$ is as follows

$$E_k = E_{\text{drm}} + E_m$$
$$= \frac{1}{4} m_1 \left( 2v_{dz}^2 + 2v_{dx}^2 + 2v_{dy}^2 + v_{tz}^2 + 2v_{t\perp}^2 + 2v_{dr}^2 \right)$$
$$+ m_1 \frac{\sqrt{\pi} v_{dr} v_{t\perp} \text{erfc}\left(-\frac{v_{dr}}{v_{t\perp}}\right)}{4\left(\exp\left(-\frac{v_{dr}^2}{v_{t\perp}^2}\right) + \frac{\pi^{1/2} v_{dr}}{v_{t\perp}} \text{erfc}\left(-\frac{v_{dr}}{v_{t\perp}}\right)\right)} + \frac{1}{4} m_2 (3v_t^2), \quad (6)$$

where $E_{\text{drm}}$ represents the energy of reactant 1 with the drift-ring-beam distribution, $E_m$ represents the energy of reactant 2 with the Maxwellian distribution, $m_1$ and $m_2$ denote the mass of the two reactants, $v_t$ is the thermal velocity of reactant 2. For the slowing-down distribution, $E_k$ is given by

$$E_k = E_{\text{sd}} + E_m = \frac{m_1}{4\ln[1+v_b^3/v_c^3]} \left\{ -v_c^2 \ln(v_c^2 - v_c v_b + v_b^2) + 2v_c^2 \ln(v_c + v_b) - 2\sqrt{3} v_c^2 \arctan\left(\frac{2v_b - v_c}{\sqrt{3} v_c}\right) + 3v_b^2 + 2\sqrt{3} v_c^2 \arctan\left(\frac{-v_c}{\sqrt{3} v_c}\right) \right\} + \frac{1}{4} m_2 (3v_t^2), \quad (7)$$

where $E_{\text{sd}}$ represents the energy of reactant 1 with the slowing-down distribution. For the kappa super-thermal distribution, $E_k$ is as follows

$$E_k = E_{\text{kappa}} + E_m = \frac{m_1 \kappa v_{th}^2}{\pi^{1/2}} \frac{\Gamma(5/2)\Gamma(\kappa-3/2)}{\Gamma(\kappa-1/2)} + \frac{1}{4} m_2 (3v_t^2), \quad (8)$$

where $E_{\text{kappa}}$ represents the energy of reactant 1 with the kappa super-thermal distribution, $\Gamma(x)$ is the gamma function.

Secondly, when the total energy $E_{\text{total}}$ remains constant, the fusion reactivity is calculated under the conditions of thermonuclear fusion, where the two reactants are Maxwellian distributions and have equal temperatures. The corresponding equation is provided as follows[7,16]

$$\langle \sigma v \rangle_M = \sqrt{\frac{8}{\pi m_r}} \frac{1}{(k_B T_r)^{3/2}} \int_0^\infty \sigma(E) E \exp\left(-\frac{E}{k_B T_r}\right) dE, \quad (9)$$

where $\langle \sigma v \rangle_M$ represents the fusion reactivity with two Maxwellian reactants of equal temperature, $E = 0.5 m_r v^2$, $v = |\boldsymbol{v}| = |\boldsymbol{v}_1 - \boldsymbol{v}_2|$, $m_r = m_1 m_2/(m_1 + m_2)$ represents reduced mass of the system, $k_B$ is the Boltzmann constant, $T_r = (m_1 T_2 + m_2 T_1)/(m_1 + m_2)$.

Thirdly, the fusion reactivity $\langle \sigma v \rangle$ is calculated using the method 1 of reference[24].

Finally, the fusion reactivity enhancement factor $f_{\langle \sigma v \rangle} = \langle \sigma v \rangle / \langle \sigma v \rangle_M$ is calculated.

Because drift-ring-beam, slowing-down, and kappa super-thermal distributions belong to different independent modules in our program, they need to be verified separately. To validate the model on the drift-ring-beam distribution, we compare it's results with previous results[7]. Figure 1 displays a comparison of the drift-ring-beam distribution results obtained in this study with previous results, which are basically consistent. It should be noted that $T_r^* = E_{\text{total}}/(3k_B)$ in Figure 1 represents the equivalent kinetic temperature, which will be referred to as "temperature" throughout the rest of the text. To validate the model on the slowing-down and kappa super-thermal distributions, we use Method 1 as described in reference[24] (currently employed in this work) and Method 3, which is weight-based method and efficiently computes the integral of equation (1) for arbitrary reactant velocity distributions, as described in reference[24] to mutually validate each other. It should be noted that these two methods are independent of each other. We present the fusion reactivity of the two distributions obtained through different Monte Carlo methods in Figure 2, showing that the results from both methods are in agreement. This verifies the program's reliability, allowing us to undertake a systematic analysis on effect of the drift-ring-beam, slowing-down, and kappa super-thermal distributions of the reactants on the fusion reactivity.

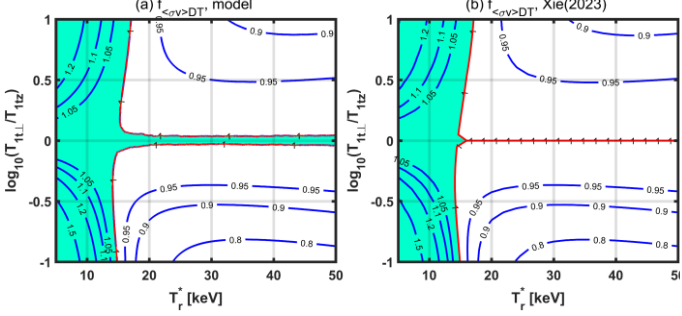

Figure 1 Drift-ring-beam distribution: D-T enhancement factor $f_{\langle\sigma v\rangle}$ of (a) model and (b) reference[7] as a function of temperature anisotropies ($T_{1t\perp}/T_{1tz}$, where $k_B T_{1t\perp} = \frac{1}{2}m_1 v_{t\perp}^2$, $k_B T_{1tz} = \frac{1}{2}m_1 v_{tz}^2$) for different temperatures $k_B T_r^* = 1/3\, E_{total}$. Other parameters are set to $T_{1dr}/T_{1t\perp} = T_{1dz}/T_{1t\perp} = 0$, $T_{1tz}/T_{1t\perp} = T_{2tz}/T_{2t\perp}$, and $T_{2t\perp}/T_{1t\perp} = 1$ with 1 and 2 representing reactant species, respectively, where $k_B T_{1dz} = \frac{1}{2}m_1 v_{1dz}^2$, $k_B T_{1dr} = \frac{1}{2}m_1 v_{1dr}^2$, $k_B T_{2t\perp} = \frac{1}{2}m_2 v_{2t\perp}^2$, $k_B T_{2tz} = \frac{1}{2}m_2 v_{2tz}^2$. The shading in the figure indicates regions where $f_{\langle\sigma v\rangle} > 1$, implying that the fusion reactivity is enhanced.

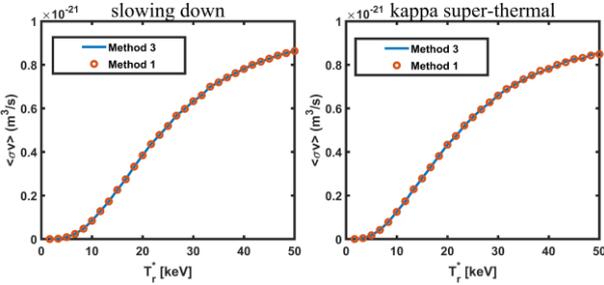

Figure 2 D-T enhancement factor $f_{\langle\sigma v\rangle}$ of (a) slowing-down distribution and (b) kappa super-thermal distribution calculated by different Monte Carlo methods as a function of temperatures $T_r^* = 1/3\, E_{total}$. The red hollow circles represent the Method 1 results and the blue lines represent the Method 3 results. For the slowing-down distribution other parameters are set to $T_{2t\perp}/T_{1c} = 1$, $T_{2tz}/T_{2t\perp} = 1$, and $T_{1b}/T_{1c} = 0.01$, where $k_B T_{1c} = \frac{1}{2}m_1 v_{1c}^2$, and $k_B T_{1b} = \frac{1}{2}m_1 v_{1b}^2$. For the kappa super-thermal distribution other parameters are set to $T_{2t\perp}/T_{1th} = 1$, $T_{2tz}/T_{2t\perp} = 1$, and $\kappa = 10$, where $k_B T_{1th} = \frac{1}{2}m_1 v_{1th}^2$.

## 3. Results

Fusion reactions involving D-T, D-D, D-He3 (Helium), and p-B11 are of great interest due to their relatively large cross sections. The D-T fusion is the most favorable for practical applications. However, with the progress of fusion research, other reactions, such as the p-B11 fusion reaction, which does not produce neutrons, are gaining attention. Recent studies on LHD have produced and observed p-B11 fusion[25], which is significant for the development of p-B11 fusion as a potential energy source and for research on alpha particles. In this study, we focus on simulating the effect of three distributions (drift-ring-beam, slowing-down, and kappa super-thermal) on both D-T and p-B11 fusion reactions, with reactant 2 (T and B11) assumed to follow a Maxwellian distribution, while the distribution of reactant 1 (D and p) is one of the three aforementioned distributions. We base our parameter selection on experimental factors and exclude unimportant parameters, such as $v_{dx}$ and $v_{dy}$. Cross-sectional data for D-T and p-B11 are obtained from references[26] and [16], respectively.

### 3.1 Drift-ring-beam distribution

The drift-ring-beam distribution is a commonly observed distribution of reactants during neutral beam injection and has been extensively studied in theoretical and simulation studies[22]. However, despite its prevalence, little research has been conducted on its effect on fusion reactivity, particularly in terms of the effect of the perpendicular ring beam velocity in the distribution. Therefore, in this study, we aim to comprehensively examine the impact of the key parameters of the drift-ring-beam distribution on the fusion reactivity, with a focus on the previously unexplored effect of the perpendicular ring beam velocity.

#### 3.1.1 Temperature anisotropy

Some simulation research has been conducted regarding the impact of temperature anisotropy on fusion reactivity[3-5,7]. It is commonly accepted that fusion reactions can be enhanced by temperature anisotropy at low temperatures. However, if critical temperatures, which are known to be 15keV( D-T ), 50keV( D-D ), 60keV( D-$^3$He ), and 140keV( p-B11 ), are exceeded, temperature anisotropy becomes counterproductive to the fusion reaction. In addition, experimental evidence is presented regarding the modification of fusion yield due to temperature anisotropy[2]. Here, we further investigated the effect of temperature anisotropy on the reaction reactivities of D-T and p-B11 fusion at different perpendicular ring beam velocities. The simulation results are presented in Figures 3 and 4, where the parameters used in the simulations are $T_{1dz}/T_{1t\perp} = 0$, $T_{1t\parallel}/T_{1t\perp} = T_{2t\parallel}/T_{2t\perp}$, and $T_{2t\perp}/T_{1t\perp} = 1$. From Figures 3(a) and (b), we can see that the critical temperatures for D-T and p-B11 fusion reactants are consistent with previous references[3-7] at $T_{1dr}/T_{1t\perp} = 0$. However, contrary to earlier results that show temperature anisotropy is unfavorable for the fusion reaction at high temperatures, there is a slight enhancement in the D-T fusion reaction at higher temperatures. The reasons for the temperature anisotropy that leads to enhance fusion reactivity at low temperatures are well understood and can be attributed to two main factors. Firstly, the anisotropy at low temperatures concentrates energy in certain directions, naturally increasing the population of the reactants at the highest-energy parts of the tails in those direction[4]. Secondly, as anisotropy increases, reactants are constrained to lower dimensions[4], which increases the probability of head-on collisions ($180°$)[5]. The physical image of the fusion reactivity enhancement, as depicted in Figure 3(a), appears differently at high temperatures as compared to low temperatures. The focus is on the increased probability of collisions happening at angles close to $0°$, which increases the probability of relative velocities being at the peak of the fusion cross section.

After considering the perpendicular ring beam velocity, as depicted in Figures 3 and 4, the fusion reactivity enhancements in D-T and p-B11 fusion significantly decrease from isotropic to anisotropic directions at low temperatures, while the fusion reactivity enhancements increase rapidly from isotropic to anisotropic directions at high temperatures. Ultimately, only enhancements at high temperatures can be observed at larger perpendicular ring beam velocities ($T_{1dr}/T_{1t\perp} \geq 9$), with corresponding critical temperatures of 50keV (D-T) and 250keV (p-B11). The above results are easily comprehensible as the perpendicular ring beam velocity concentrates energy in the perpendicular direction of reactant 1, leading to decreasing the population of the reactants at the highest-energy parts of the tails at low temperatures. Furthermore, the emergence of the critical temperature indicates that the effect of temperature anisotropy near the critical temperature is insignificant at $T_{1d\perp}/T_{1t\perp} \geq 9$ because, at this point, energy is mainly concentrated in drift energy. After exceeding the critical temperature, the total energy increases, and the corresponding thermal energy also increases, thus the effect of temperature anisotropy reappears. Finally, the main conclusions regarding D-T and p-B11 in the temperature range we focus on, namely, 5-50 keV (D-T) and 100-500 keV (p-B11), are as follows:

$T_{1dr}/T_{1t\perp} = 0$: Temperature anisotropy is advantageous in the low temperature range, disadvantageous in the moderate temperature range, and advantageous in the high temperature range;

$T_{1dr}/T_{1t\perp} > 0$: The enhancement decreases in the low temperature range and increases in the high temperature range. More specifically, there is an overlap between the key temperature range of D-T and the low temperature range, while D-T fusion reactivity enhancement is about -90% to -10%. There is also an overlap between the key temperature range of p-B11 and the high temperature range, while p-B11 fusion reactivity enhancement is about 5% to 200%.

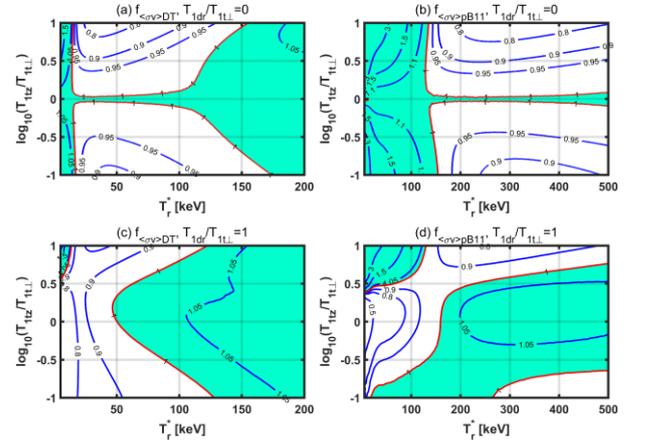

Figure 3 Enhancement factor $f_{\langle\sigma v\rangle}$ of (a) D-T: $T_{1dr}/T_{1t\perp} = 0$, (b) p-B11: $T_{1dr}/T_{1t\perp} = 0$, (c) D-T: $T_{1dr}/T_{1t\perp} = 1$, and (d) p-B11: $T_{1dr}/T_{1t\perp} = 1$, as a function of temperature anisotropies ($T_{1tz}/T_{1t\perp}$) for different temperatures $k_B T_r^* = 1/3\, E_{\text{total}}$.

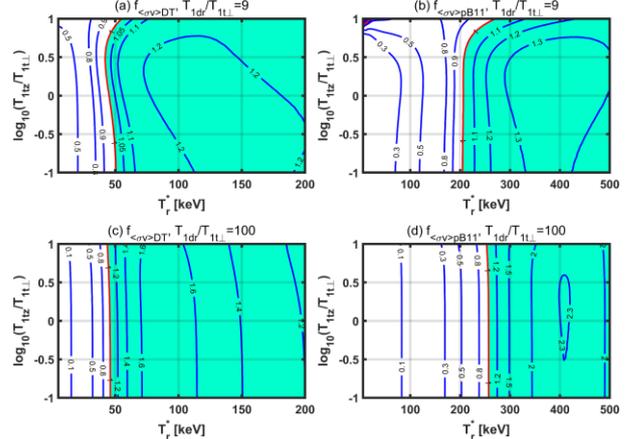

Figure 4 Enhancement factor $f_{\langle\sigma v\rangle}$ of (a) D-T: $T_{1dr}/T_{1t\perp} = 9$, (b) p-B11: $T_{1dr}/T_{1t\perp} = 9$, (c) D-T: $T_{1dr}/T_{1t\perp} = 100$, and (d) p-B11: $T_{1dr}/T_{1t\perp} = 100$, as a function of temperature anisotropies ($T_{1tz}/T_{1t\perp}$) for different temperatures $k_B T_r^* = 1/3\, E_{\text{total}}$.

*3.1.2 Unequal reactant temperatures*

When the total energy remains constant, a higher temperature for the lighter reactants compared to the heavier reactants is advantageous for the fusion reaction, and the higher the temperature of the lighter reactants, the greater the enhancement in the fusion reaction[13]. Our simulation results, as shown in Figures 5(a) and (b) (showing the effect of unequal reactant temperatures on the fusion reactivity enhancement factor $f_{\langle\sigma v\rangle}$ at different perpendicular ring beam velocity), support this view. However, our simulation results indicate that once the temperature exceeds 60 keV, heavier reactants having higher temperatures are more advantageous for D-T fusion reaction, as depicted in Figure 5(a). The above results are easily comprehensible as unequal temperatures of two reactants result in energy being concentrated in the reactant with higher temperature. At low temperatures, energy is concentrated in the lighter reactant (i.e., reactant 1), leading to two favorable factors for fusion reactions. One is an increase in the population of reactant 1 at the highest-energy parts of the tails, and the other is the relative velocity between the two reactants being close to the velocity of reactant 1.

When the perpendicular ring beam velocity is considering, the fusion reactivity enhancement factor $f_{\langle\sigma v\rangle}$ decreases in the low temperature range and increases in the high temperature range. This is similar to the phenomenon of temperature anisotropy and is governed by the same physical mechanism. However, a comparison between Figure 4 (b) with Figure 6 (b) indicates that the effect of unequal reactant temperatures at higher perpendicular ring beam velocities remains significant in p-B11 fusion reaction. Finally, the main conclusions regarding D-T and p-B11 in the temperature range we focus on, namely, 5-50 keV (D-T) and 100-500 keV (p-B11), are as follows:

$T_{1dr}/T_{1t\perp} = 0$: The reactants D and p with higher temperatures are advantageous for fusion reactions;

$T_{1dr}/T_{1t\perp} > 0$: The low temperature range overlaps with the key temperature range of D-T, and D-T fusion reactivity enhancement is about -90% to -10%; the high temperature range overlaps with the key temperature range of p-B11, and p-B11 fusion reactivity enhancement is about 5% to 200%.

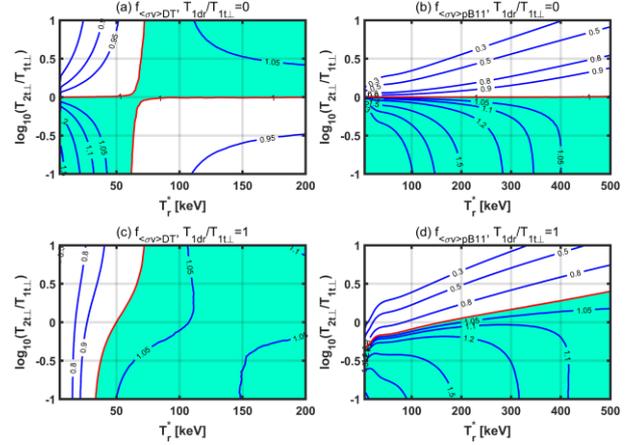

Figure 5 Enhancement factor $f_{\langle\sigma v\rangle}$ of (a) D-T: $T_{1dr}/T_{1t\perp} = 0$, (b) p-B11: $T_{1dr}/T_{1t\perp} = 0$, (c) D-T: $T_{1dr}/T_{1t\perp} = 1$, and (d) p-B11: $T_{1dr}/T_{1t\perp} = 1$, as a function of unequal reactant temperatures ($T_{2t\perp}/T_{1t\perp}$) for different temperatures $k_B T_r^* = 1/3\, E_{total}$. Other parameters are set to $T_{1dz}/T_{1t\perp} = 0$, and $T_{1t\parallel}/T_{1t\perp} = T_{2t\parallel}/T_{2t\perp} = 1$.

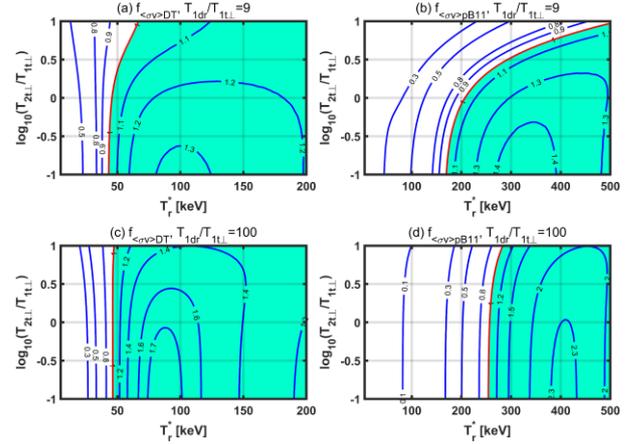

Figure 6 Enhancement factor $f_{\langle\sigma v\rangle}$ of (a) D-T: $T_{1dr}/T_{1t\perp} = 9$, (b) p-B11: $T_{1dr}/T_{1t\perp} = 9$, (c) D-T: $T_{1dr}/T_{1t\perp} = 100$, and (d) p-B11: $T_{1dr}/T_{1t\perp} = 100$, as a function of unequal reactant temperatures ($T_{2t\perp}/T_{1t\perp}$) for different temperatures $k_B T_r^* = 1/3\, E_{total}$. Other parameters are set to $T_{1dz}/T_{1t\perp} = 0$, and $T_{1t\parallel}/T_{1t\perp} = T_{2t\parallel}/T_{2t\perp} = 1$.

### 3.1.3 Drift velocity in parallel direction

When the drift velocity in parallel direction, i.e., $v_{d\parallel}$, is the only factor considered, the enhancement factor in a specific parameter range of p-B11 fusion increases by about 20%, significantly relaxing the Lawson criterion in p-B11 fusion[7]. When only the perpendicular ring beam velocity $v_{dr}$ is taken into account, the results of the previous sections indicate that the perpendicular ring beam velocity reduces the enhancement factor in the low temperature range but increases it in the high temperature range. However, in experiments, both of them often coexist, particularly under conditions such as neutral beam injection. Therefore, it is natural to study the combined effect of both on the fusion reactions. The other parameters are set to $T_{2t\perp}/T_{1t\perp} = 1$, and $T_{1tz}/T_{1t\perp} = T_{2tz}/T_{2t\perp} = 1$. Figures 7 and 8 illustrate the effect of drift velocity in parallel direction on fusion reactivity enhancement factor $f_{\langle\sigma v\rangle}$ at different perpendicular ring beam velocities. As depicted in Figures 7(a) and (b), the drift velocity in parallel direction reduces the fusion reactivity at low temperatures but enhances it at high temperatures in the absence of perpendicular ring beam velocity, which is similar to the perpendicular ring beam velocity. From Figures 7(c), (d), and 8, it can be observed that when both are present, the enhancement factors in both the low and high temperature ranges rapidly decrease and increase, respectively, and the critical temperature appears at lower perpendicular ring beam velocity. In the process mentioned above, the drift velocity in parallel direction plays a role in further accelerating the concentration of energy in reactant 1. Finally, the main conclusions regarding D-T and p-B11 in the temperature range we focus on, namely, 5-50 keV (D-T) and 100-500 keV (p-B11), are as follows:

$T_{1dr}/T_{1t\perp} \geq 0$: The effect of the drift velocity in parallel direction on fusion reactions is consistent with that of the perpendicular ring beam velocity. Specifically, D-T fusion reactivity enhancement is about -90% to -5%; p-B11 fusion reactivity enhancement is about 10% to 200%.

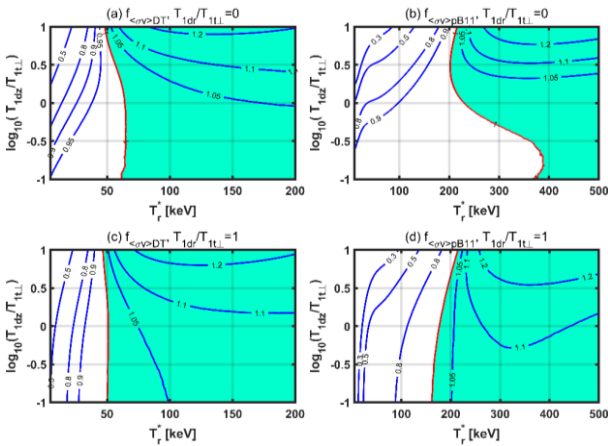

Figure 7 Enhancement factor $f_{\langle\sigma v\rangle}$ of (a) D-T: $T_{1dr}/T_{1t\perp} = 0$, (b) p-B11: $T_{1dr}/T_{1t\perp} = 0$, (c) D-T: $T_{1dr}/T_{1t\perp} = 1$, and (d) p-B11: $T_{1dr}/T_{1t\perp} = 1$, as a function of drift velocity in parallel direction ($T_{1dz}/T_{1t\perp}$) for different temperatures $k_B T_r^* = 1/3\, E_{total}$.

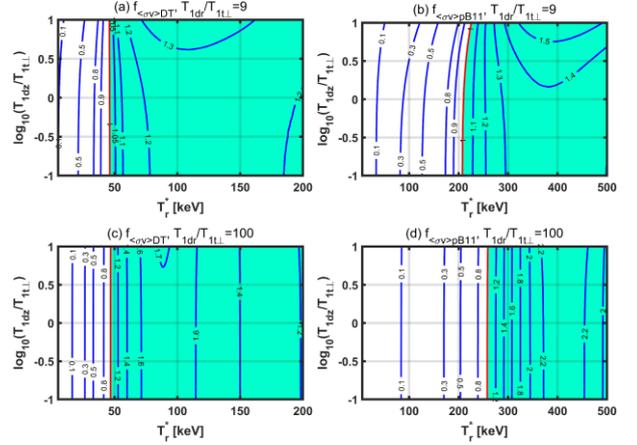

Figure 8 Enhancement factor $f_{\langle\sigma v\rangle}$ of (a) D-T: $T_{1dr}/T_{1t\perp} = 9$, (b) p-B11: $T_{1dr}/T_{1t\perp} = 9$, (c) D-T: $T_{1dr}/T_{1t\perp} = 100$, and (d) p-B11: $T_{1dr}/T_{1t\perp} = 100$, as a function of drift velocity in parallel direction ($T_{1dz}/T_{1t\perp}$) for different temperatures $k_B T_r^* = 1/3\, E_{total}$.

### 3.1.4 Reactant 1 with temperature anisotropy

In section 3.1.1, we considered the case where both reactants have temperature anisotropy, but this specific case is not universal. Thus, in this subsection we investigate the relatively common case where reactant 1 has temperature anisotropy while reactant 2 follows the Maxwellian distribution. The other parameters are set to $T_{1dz}/T_{1t\perp} = 0$, $T_{2t\perp}/T_{1t\perp} = 1$, and $T_{2tz}/T_{2t\perp} = 1$. Figures 9 and 10 show the effect of the temperature anisotropy of reactant 1 on the fusion reactivity enhancement factor $f_{\langle\sigma v\rangle}$ at different perpendicular ring beam velocities. A comparison between Figure 3 with Figure 9 indicates that when we only consider reactant 1 to be anisotropic instead of both, there are significant changes in the enhancement interval of the D-T and p-B11 fusion reactions, especially at $T_{1dr}/T_{1t\perp} = 0$, where the enhancement interval on temperatures is expanded and the enhancement is intensified. This phenomenon can be explained by the fact that the dimension in which the reactant is constrained changes depending on whether one or both reactants display temperature anisotropy. Once we take into account the perpendicular ring beam velocity, as shown in Figures 9(c), (d), and 10, the enhancement interval of D-T fusion reaction is quickly dominated by it, while the enhancement interval of P-B fusion reaction still exists at low temperatures, particularly at a large perpendicular drift velocity ($T_{1dr}/T_{1t\perp} = 9$). After further increasing the perpendicular ring beam velocity, as shown in figures 10(c) and (d), a critical temperature appears, and the entire high-temperature range exhibits enhancement. Finally, the main conclusions regarding D-T and p-B11 in the temperature range we focus on, namely, 5-50 keV (D-T) and 100-500 keV (p-B11), are as follows:

$T_{1dr}/T_{1t\perp} = 0$ : The enhancement interval on temperatures is expanded;

$T_{1dr}/T_{1t\perp} > 0$: D-T fusion reactivity enhancement is about -90% to 50%, and p-B11 fusion reactivity enhancement is about 10% to 200%.

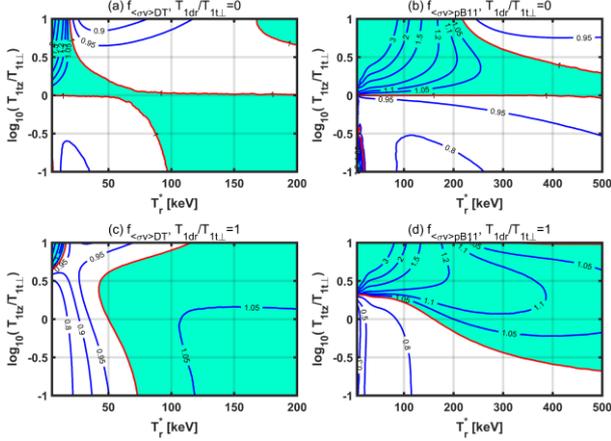

Figure 9 Enhancement factor $f_{\langle\sigma v\rangle}$ of (a) D-T : $T_{1dr}/T_{1t\perp} = 0$, (b) p-B11: $T_{1dr}/T_{1t\perp} = 0$, (c) D-T : $T_{1dr}/T_{1t\perp} = 1$, and (d) p-B11: $T_{1dr}/T_{1t\perp} = 1$, as a function of temperature anisotropy (reactant 1) for different temperatures $k_B T_r^* = 1/3\, E_{total}$.

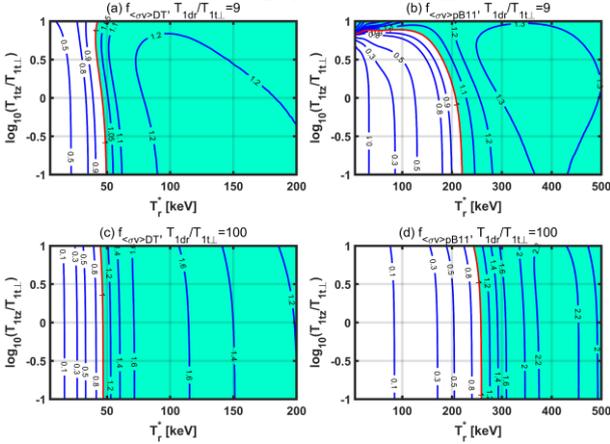

Figure 10 Enhancement factor $f_{\langle\sigma v\rangle}$ of (a) D-T : $T_{1dr}/T_{1t\perp} = 9$, (b) p-B11: $T_{1dr}/T_{1t\perp} = 9$, (c) D-T : $T_{1dr}/T_{1t\perp} = 100$, and (d) p-B11: $T_{1dr}/T_{1t\perp} = 100$, as a function of temperature anisotropy (reactant 1) for different temperatures $k_B T_r^* = 1/3\, E_{total}$.

### 3.1.5 Perpendicular ring beam velocity

The research on perpendicular ring beam velocity in the previous sections only involved several specific values. In this section, we investigate in greater detail the effect of perpendicular ring beam velocity on the D-T and p-B11 fusion reactivities, as shown in Figure 11 where parameters are set to $T_{1tz}/T_{1t\perp} = T_{2t\perp}/T_{1t\perp} = T_{2tz}/T_{2t\perp} = 1$, and $T_{1dz}/T_{1t\perp} = 0$. From the figure, we can see that D-T fusion reaction is highly sensitive to perpendicular ring beam velocity, and reaches critical temperature at very low perpendicular ring beam velocities. In contrast, p-B11 fusion reaction reaches critical temperature only at higher perpendicular ring beam velocities.

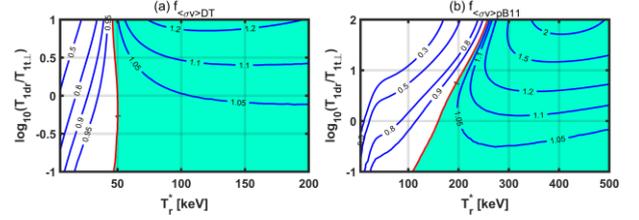

Figure 11 Enhancement factor $f_{\langle\sigma v\rangle}$ of (a) D-T, and (b) p-B11, as a function of perpendicular ring beam velocity $(T_{1dr}/T_{1t\perp})$ for different temperatures $k_B T_r^* = 1/3\, E_{total}$.

### 3.2 Slowing-down distribution

After neutral beam injection, the distribution of reactants can be well-described by the drift-ring-beam distribution. However, over time and as a result of collisions and other interactions, this distribution tends to shift towards the slowing-down distribution[22]. As a key component of the continuous evolution of the aforementioned reactant distribution, more work on the slowing-down distribution is necessary.

#### 3.2.1 Unequal reactant temperatures

Since we only take into account the isotropic slowing-down distribution, the parameters available for research are limited to the unequal reactant temperatures and birth speed. Here, we focused on studying the effect of unequal reactant temperatures on the fusion reactivity enhancement factor $f_{\langle\sigma v\rangle}$ at different birth speeds, as shown in Figures 12 and 13. The other parameters are set to $T_{2tz}/T_{2t\perp} = 1$. It can be observed from both figures that as the birth speed increases, the enhancement range of both D-T and p-B11 fusion reactions shifted towards lower temperatures, while the magnitude of the enhancement increases. Moreover, when $T_{1b}/T_{1c} \leq 9$, the lighter reactants having higher temperatures at low temperatures are more advantageous for the fusion reaction, while the heavier reactants having higher temperatures at high temperatures are more advantageous for the fusion reaction. As the birth speeds continue to increase, it becomes the dominant factor, and the effect of unequal reactant temperatures gradually diminished. In addition, the effect of the birth speed on p-B11 fusion reaction is even more significant when the $T_{1b}/T_{1c} = 100$, as evidenced by an enhancement factor $f_{\langle\sigma v\rangle}$ exceeding 10 in the low-temperature range. Figure 14, which shows the slowing-down distribution function at different birth speeds when $E_{total} = 5\text{keV}$, provides us with the answer on the above phenomenon, which can be attributed to the significant increase in the population of reactant 1 at the highest-energy parts of the tails. Combining the enhancement intervals of the drift-ring-beam distribution and the slowing-down distribution, we seem to be able to find a specific parameter range in which the fusion reactions continue to enhancement after the neutral beam injection. Finally, the main conclusions regarding D-T and p-B11 in the temperature range we focus on, namely, 5-50 keV (D-T) and 100-500 keV (p-B11), are as follows:

$1 < T_{1b}/T_{1c} \leq 9$: The enhancement increases in the low temperature range and decreases in the high temperature range. Specifically, D-T fusion reactivity enhancement is about 10% to 25%; p-B11 fusion reactivity enhancement is about 20% to 300%;

$T_{1b}/T_{1c} = 100$: The enhancement increases significantly at low temperatures, especially for p-B11 fusion reaction, where the enhancement factor can be larger than 10.

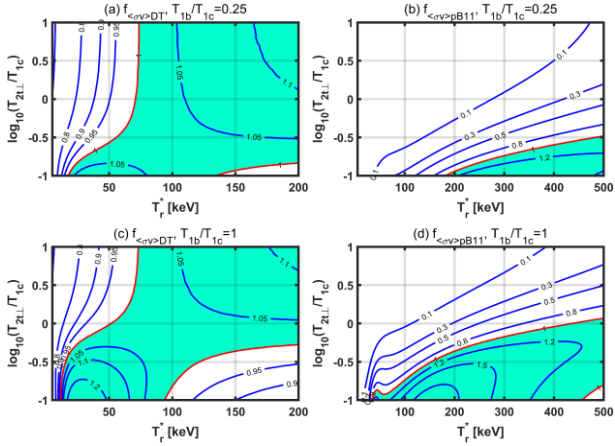

Figure 12 Enhancement factor $f_{\langle\sigma v\rangle}$ of (a) D-T: $T_{1b}/T_{1c} = 0.25$, (b) p-B11: $T_{1b}/T_{1c} = 0.25$, (c) D-T: $T_{1b}/T_{1c} = 1$, and (d) p-B11: $T_{1b}/T_{1c} = 1$, as a function of unequal reactant temperatures ($T_{2t\perp}/T_{1c}$) for different temperatures $k_B T_r^* = 1/3\, E_{total}$.

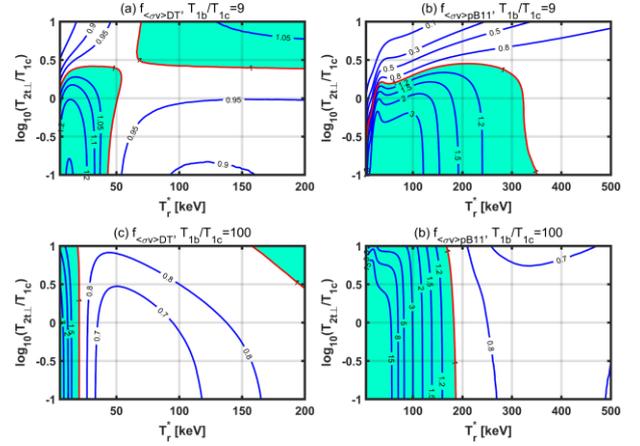

Figure 13 Enhancement factor $f_{\langle\sigma v\rangle}$ of (a) D-T: $T_{1b}/T_{1c} = 9$, (b) p-B11: $T_{1b}/T_{1c} = 9$, (c) D-T: $T_{1b}/T_{1c} = 100$, and (d) p-B11: $T_{1b}/T_{1c} = 100$, as a function of unequal reactant temperatures ($T_{2t\perp}/T_{1c}$) for different temperatures $k_B T_r^* = 1/3\, E_{total}$.

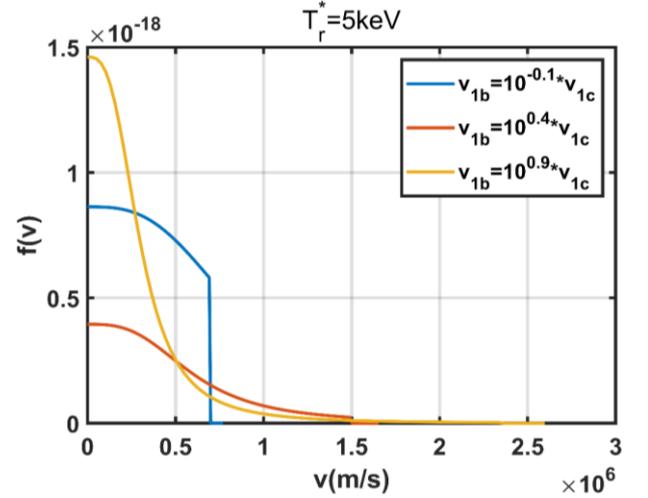

Figure 14 Slowing-down distribution with different birth speeds when $k_B T_r^* = 5\text{keV}$, $T_{2tz}/T_{2t\perp} = 1$.

### 3.2.2 Birth speed

In this section, we present a more detailed results on the birth speed. Figure 15 shows the effect of birth speed on the fusion reactivity enhancement factor $f_{\langle\sigma v\rangle}$ for both reactions. As shown in the figure, an enhancement factor higher than one can only be achieved with a high birth speed at low temperature ranges for both reactions, while at high temperature range, a relatively lower birth speed is required.

**9 / 12**

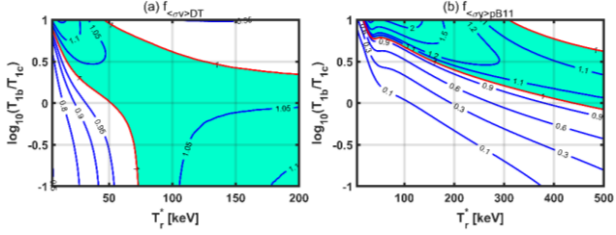

Figure 15 Enhancement factor $f_{\langle\sigma v\rangle}$ of (a) D-T, and (b) p-B11, as a function of birth speed ($T_{1b}/T_{1c}$) for different temperatures $k_B T_r^* = 1/3\, E_{total}$.

### 3.3 Kappa super-thermal distribution

In magnetic confinement fusion plasmas, the velocity distribution of reactants can form a high-energy tail under certain conditions, such as the ion cyclotron resonance heating (ICRH). Previous theoretical studies have suggested that the high-energy tail is present during the fusion reaction and affects the fusion reactivity[10]. Therefore, it is of great significance to study the effect of reactants with a high-energy tail on fusion reactivity during the fusion reaction. The kappa super-thermal distribution[23] is a useful model in describing this high-energy tail. Previous references[12,13] on the kappa super-thermal distribution has studied the effect of both the supra-thermal and thermal components of this distribution on fusion reactivity. In this section, we have conducted a detailed study on the isotropic kappa super-thermal distribution, focusing on key parameters of the unequal reactant temperatures and high energy power-law index $\kappa$.

#### 3.3.1 Unequal reactant temperatures

The distinctive feature of the kappa super-thermal distribution is its similarity to the Maxwellian distribution for low energies while exhibiting a pronounced tail for high energies. As $\kappa$ approaches infinity, the distribution approaches the Maxwellian distribution. Therefore, the effect of kappa super-thermal distribution on the fusion reactivity under medium and low $\kappa$ conditions is what we are interested in. From the analysis of the drift-ring-beam distribution and slowing-down distribution, when $\kappa$ is small, the high-energy tail of the kappa super-thermal distribution is advantageous in the low temperature range. Hence, the results presented in Figures 16(a) and (b) are expected. As $\kappa$ increases, the trend that a higher temperature for the lighter reactants at low temperatures is favorable for the fusion reactions, while a higher temperature for heavier reactants at high temperatures is favorable, gradually emerges. Overall, when $\kappa \geq 10$, the enhancement range of the kappa super-thermal distribution significantly overlaps with the enhancement range resulting from the unequal temperatures of two reactants with Maxwellian distribution (as shown in Figures 5 (a) and (b)), and the former's enhancement is lower than the latter. From this perspective, the kappa super-thermal distribution only provides a meaningful enhancement when $\kappa$ is small. Finally, the main conclusions regarding D-T and p-B11 in the temperature range we focus on, namely, 5-50 keV (D-T) and 100-500 keV (p-B11), are as follows:

$\kappa = 2$: The low temperature enhancement range partially overlaps with the key temperature range for D-T and P-B11, with enhancements of 5%-200% for D-T and 10%-50% for P-B11.

$\kappa > 10$: As $\kappa$ increases, the kappa super-thermal distribution approaches the Maxwellian distribution.

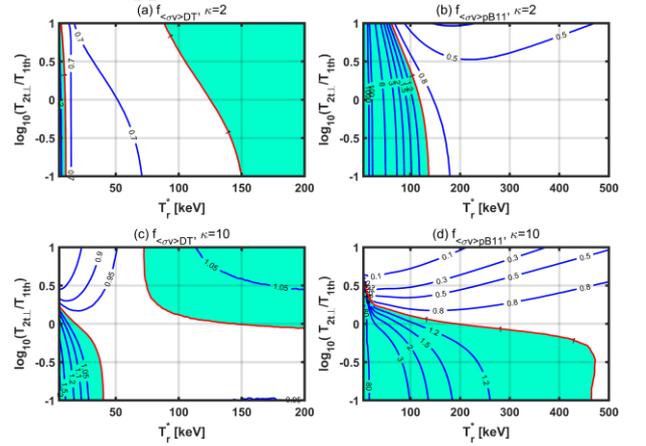

Figure 16 Enhancement factor $f_{\langle\sigma v\rangle}$ of (a) D-T: $\kappa = 2$, (b) p-B11: $\kappa = 2$, (c) D-T: $\kappa = 10$, and (d) p-B11: $\kappa = 10$, as a function of unequal reactant temperatures ($T_{2t\perp}/T_{1th}$) for different temperatures $k_B T_r^* = 1/3\, E_{total}$.

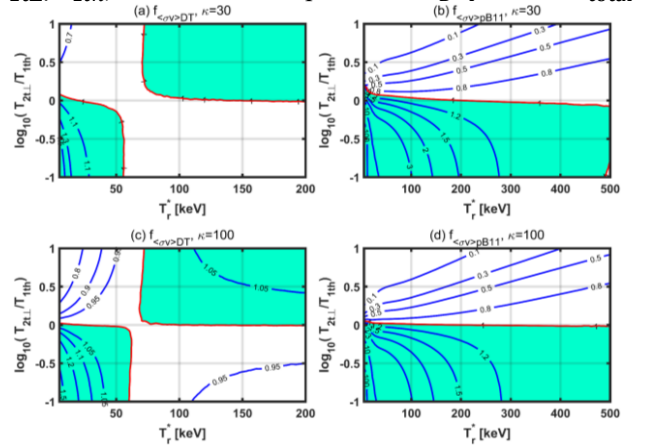

Figure 17 Enhancement factor $f_{\langle\sigma v\rangle}$ of (a) D-T: $\kappa = 30$, (b) p-B11: $\kappa = 30$, (c) D-T: $\kappa = 100$, and (d) p-B11: $\kappa = 100$, as a function of unequal reactant temperatures ($T_{2t\perp}/T_{1th}$) for different temperatures $k_B T_r^* = 1/3\, E_{total}$.

*3.3.2 Parameter $\kappa$*

In this section, we present more detailed results on the effect of $\kappa$ on fusion reactivity enhancement, as shown in Figure 18. In this simulation, we set the temperatures of the reactants to be equal, considering only the effect of $\kappa$ on fusion enhancement. As can be seen from the figure, both D-T and p-B11 fusion reactions exhibit significant enhancements in the low-temperature range, mainly attributed to the high-energy tail of the kappa super-thermal distribution. As expected, the enhancements gradually decrease as $\kappa$ increases.

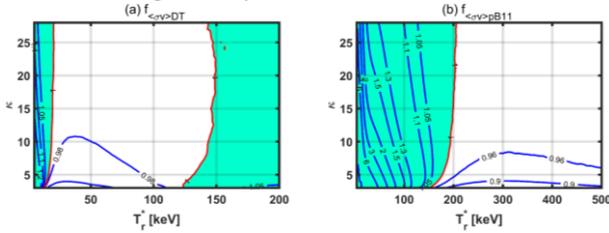

Figure 18 Enhancement factor $f_{\langle\sigma v\rangle}$ of (a) D-T, and (b) p-B11, as a function of $\kappa$ for different temperatures $k_B T_r^* = 1/3\, E_{\text{total}}$.

## 4. Conclusion

In this work, we develop a program to compute the enhancement of fusion reactivity relative to the Maxwellian distribution while keeping the total kinetic energy constant under non-Maxwellian distributions. In particular, we conduct a detailed analysis and study on the effects of drift-ring-beam, slowing-down, and kappa super-thermal distributions on the fusion reactivity enhancement for D-T and p-B11 fusion reactions.

Overall, for the temperature ranges of interest to us, namely $5\,\text{keV} < T_r^* < 50\,\text{keV}$ (D-T) and $100\,\text{keV} < T_r^* < 500\,\text{keV}$ (p-B11), all three distributions exhibit enhancement ranges. For the drift-ring-beam distribution, the enhancement factors for both reactions are affected by the perpendicular ring beam velocity, leading to decreased enhancement at low temperature range and increased enhancement at high temperature range, which is independent of the other parameters in the distribution and is also irrelevant to the type of fusion reaction. This is favorable for p-B11 fusion but not for D-T fusion. This is because the important temperature range for p-B11 fusion reaction significantly overlaps with the high temperature enhancement range, while the important temperature range for D-T fusion significantly overlaps with the low temperature reduction range. Through further study of other key parameters when the perpendicular ring beam velocity is zero, we find that the temperature anisotropy of the two reactants also leads to a slight enhancement in the high-temperature range; the effects of the temperature anisotropy of a single reactant and both reactants on D-T and p-B11 fusion reactions differs significantly and the former has a larger enhancement ranges and stronger enhancement with respect to temperature; the research on the unequal reactant temperatures shows that a higher temperature of the heavier reactant is more advantageous for fusion reactions in the high temperature range. For the slowing-down distribution, the birth speed plays a crucial role in determining the enhancement range of D-T and p-B11 fusion reactions, and increasing birth speed leads to a shift in the enhancement ranges towards lower temperatures. This is beneficial for both D-T and p-B11 fusion reactions, not only because it overlaps with the key temperature range for D-T and p-B11 fusion reactions, but also because the enhancement is larger at higher birth speeds. Specifically, when $1 < T_{1b}/T_{1c} \leq 9$, the enhancement for D-T fusion reaction is about 10% to 25%, and the enhancement for p-B11 fusion reaction is about 20% to 300%. Finally, for the kappa super-thermal distribution, it has a significant effect on the fusion reactivity enhancement at low temperatures when $\kappa$ is small.

The present work can be also extended to other distributions and other fusion reactions, such as D-D and D-He3 fusion, which could be future works.

**Acknowledgments**

This work is supported by the compact fusion project in ENN group and by the High-End Talents Program of Hebei Province, Innovative Approaches towards Development of Carbon-Free Clean Fusion Energy (No. 2021HBQZYCSB006).